\documentclass[11pt,twoside]{article}

\usepackage{asp2006}
\usepackage{epsf}
\usepackage{graphicx}
\usepackage{natbib}
\usepackage{amssymb}
\begin{document}
\title{The Milky Way Nuclear Star Cluster in Context}   
\author{Rainer Sch\"odel}   
\affil{Instituto de Astrof\'isica de Andaluc\'ia - CSIC, Glorieta de
  la Astronom\'ia s/n, 18008 Granada, Spain}

\begin{abstract} 
  Nuclear star clusters are located at the dynamical centers of the
  majority of galaxies. They are usually the densest and most massive
  star cluster in their host galaxy. In this article, I will give a
  brief overview of our current knowledge on nuclear star clusters and
  their formation. Subsequently, I will introduce the nuclear star
  cluster at the center of the Milky Way, that surrounds the massive
  black hole, Sagittarius\,A* (Sgr\,A*). This cluster is a unique
  template for understanding nuclear star clusters in general because
  it is the only one of its kind which we can resolve into individual
  stars. Thus, we can study its structure, dynamics, and population in
  detail. I will summarize our current knowledge of the Milky Way
  nuclear star cluster, discuss its relation with nuclear clusters in
  other galaxies and point out where further research is needed.
\end{abstract}

\section{Nuclear star clusters}

The study of nuclear star clusters (NSCs) of galaxies has only become
feasible from the middle of the 1990s on with the high sensitivity and
spatial resolution offered by the Hubble Space Telescope. NSCs are
detected in 50\%-80\% of spiral, (d)E, and S0 galaxies
\citep[e.g.,][]{Phillips:1996nx,Carollo:1998fk,Matthews:1999uq,Boker:2002kx,Boker:2004oq,Balcells:2003vn,Graham:2003fk,Cote:2006eu}.
It is not always possible to identify NSCs unambiguously,
though. Different galaxy types pose different obstacles to their
detection. For example, in some galaxies an NSC may be hidden by a
high central surface brightness, while in late spirals with low surface
density the exact center may be hard to determine, making it difficult
to choose between several candidate clusters. Therefore, the reported
detection rates can be regarded as lower limits to their frequency of
occurrence which may possibly reach up to 100\%. NSCs appear to be
absent in massive ellipticals however. Some of them show 
central ``extra light'', but this appears to be a phenomenon distinct from
true NSCs \citep{Kormendy:2009fk}.

\begin{figure}[!htb]
\centering
\includegraphics[width=.8\textwidth]{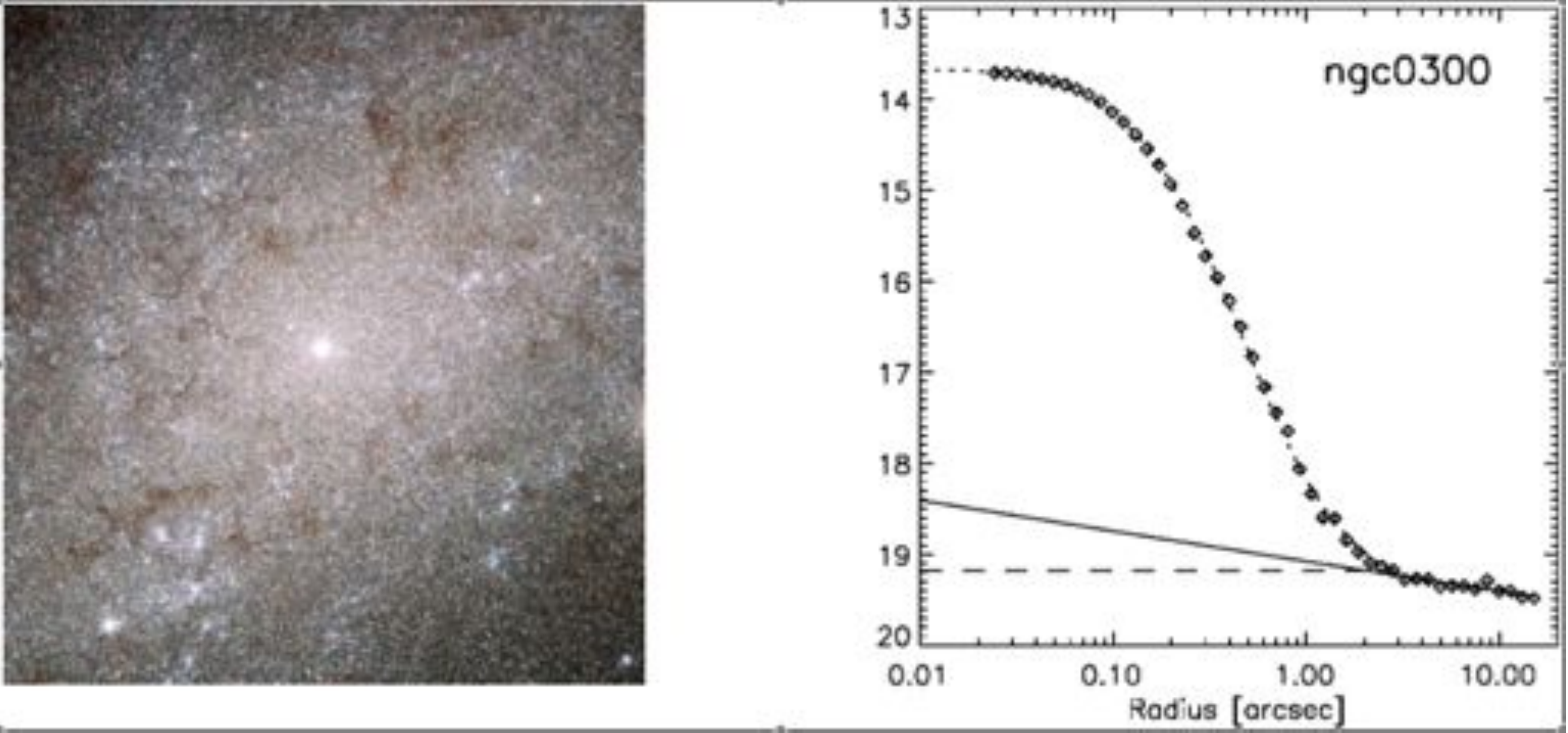}
\caption{\label{Fig:NSC} Left: HST/ACS image of the center of the Sd galaxy
  NGC\,300. The image size is
  $1.2\,\mathrm{kpc}\times1.2\,\mathrm{kpc}$. Right: $I$-band surface
  brightness profile of NGC\,300, as determined from HST/WFPC2
  data. This figure has been adapted from
  \citet{van-der-Marel:2007bh} with kind permission of the authors, see their work for details.}
\end{figure}

A concise review on the properties of NSCs is given in
\citet{Boeker:2009eu}. NSCs have typically half-light radii of
2-5\,pc and masses of $10^{6} - 10^{7}$\,M$_{\odot}$. Consequently,
they are among the most massive and certainly densest star clusters in
the Universe. The star formation histories of NSCs are complex and a
large fraction of them show evidence for several generations of stars
with the youngest ones having frequently ages $<100$\,Myr
\citep{Rossa:2006zr,Walcher:2006ve}. Hence, NSCs appear to be
characterized by repeated episodes of star formation.

Two hypotheses have been suggested for the formation of NSCs: (a)
\emph{Migratory formation} via (repeated) infall of globular clusters
or starburst-like clusters into galaxy centers through dynamical
friction. (b) \emph{In-situ formation} via repeated gas infall and
subsequent star formation. It may well be  that both mechanisms
contribute to the buildup of NSCs over a Hubble time. The discovery of
rotation of the NSCs of NGC\,4244 \citep{Seth:2008kx} and of the Milky
Way \citep{Trippe:2008it,Schodel:2009zr} combined with the 
 observation that many NSCs show signs of star formation within
the last 100 Myr suggest that they form at least partially from
infalling material from the host galaxy.

The recent realization that NSCs appear to obey similar scaling
relationships with respect to their host galaxy masses as do massive
black holes (MBHs) has raised even higher interest in these still
poorly understood objects
\citep{Rossa:2006zr,Wehner:2006gf,Ferrarese:2006ly,Balcells:2007ly,Graham:2009lh}. NSCs
have been found to coexist with massive black holes in a small but
steadily increasing (due to improved observations) number of cases
\citep[e.g.,][]{Filippenko:2003qf,Gonzalez-Delgado:2008wd,Seth:2008rr,Kormendy:2009fk}. Based
on a compilation of the cases of coexisting NSCs and MBHs with
reliably estimated masses, \citet{Graham:2009lh} show that the masses
of the nuclei of the most massive ellipticals appear to be dominated
by SMBHs, with NSCs not detected in many cases. The
nuclear masses of the least massive spheroids appear to be dominated,
however, by the masses of their respective NSCs. There is a
transitional zone between these two regimes, where MBHs and NSCs
clearly coexist \citep[see Fig.\,1 in][]{Graham:2009lh}.

Studying NSCs is a difficult task. Because of their great distances
and compact sizes, NSCs have, on average, apparent diameters
$\lesssim1"$ and apparent $I$-band magnitudes around $18-22$
\citep[e.g.,][]{Boker:2002kx}.  They are barely spatially resolved,
even with the HST or with 8-10m ground-based telescopes equipped with
adaptive optics.  This makes the nuclear cluster at the center of the
Milky Way (MW) a special target: It is the only NSC that can be resolved
down to milli-parsec scales with current instrumentation. It serves as
a unique template to study the interaction of a massive black hole
with the surrounding nuclear star cluster and interstellar matter.

\section{Large scale structure of the  Milky Way NSC}

\begin{figure}[!htb]
\centering
\includegraphics[width=\textwidth]{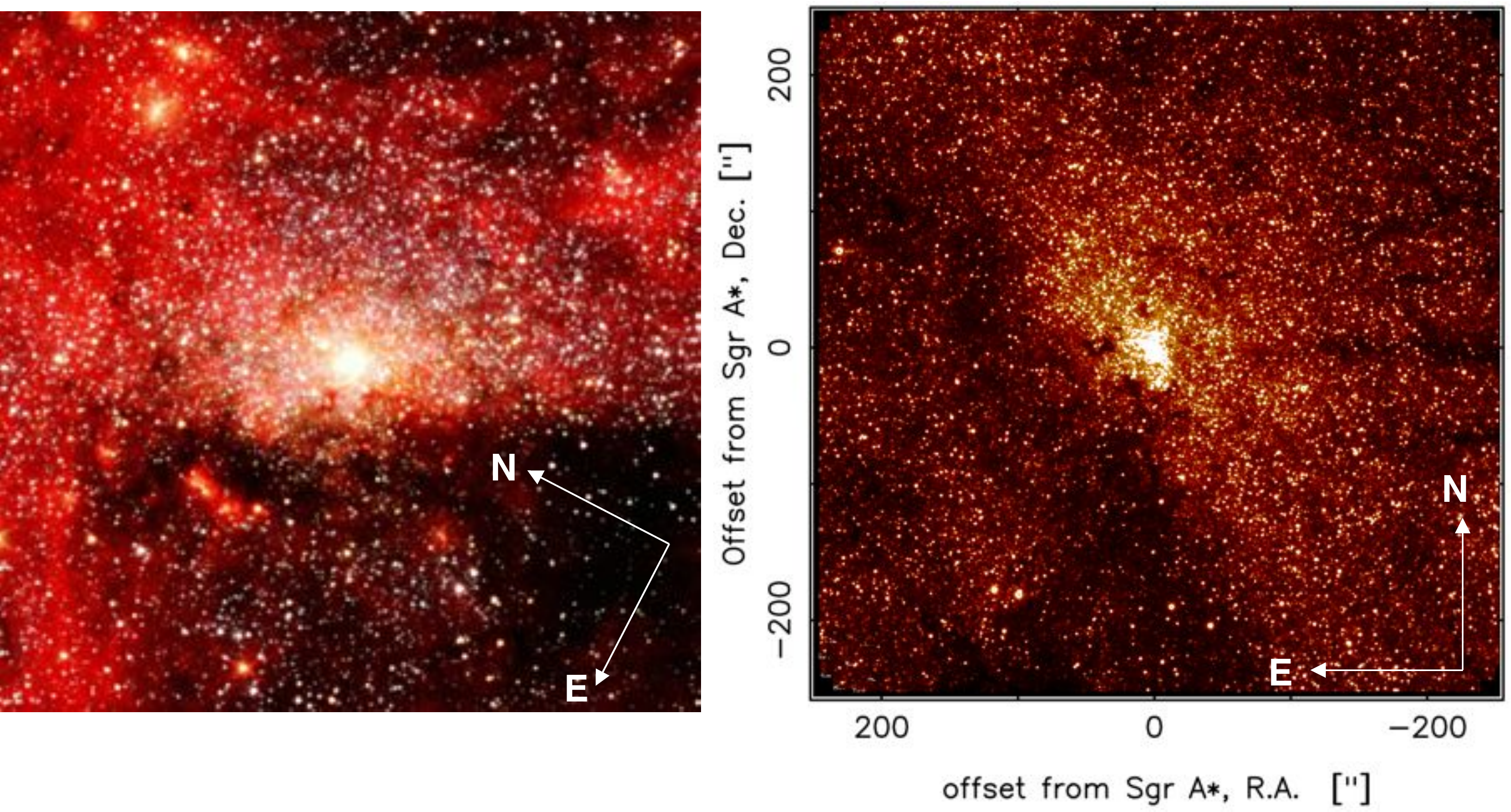}
\caption{\label{Fig:MWNSC} Left: Spitzer image of the nuclear star cluster of the Milky
  Way with IRAC, composed of observations at $3.6$, $4.5$, $5.8$, and
  $8.0$\,$\mu$m (Credit: NASA/JPL-Caltech/S. Stolovy,
  SSC/Caltech). The size of the image is approximately $10'\times 10'$
  or $20\,\mathrm{pc}\times20\,\mathrm{pc}$ Right: $Ks$-band image
  of the GC with IRSF/SIRIUS \citep[see][]{Nishiyama:2006tx,Nishiyama:2009oj}.}
\end{figure}

The nuclear star cluster at the center of the Milky Way was detected
in the pioneering NIR observations of the Galactic center by
\citet{Becklin:1968nx}. It was described as a dominant source $5'$ in
diameter, but its true nature was not clear, given the lack of
observations of extragalactic NSCs at that time. Recent images of the
MW NSC are shown in Fig.\,\ref{Fig:MWNSC}. The first publications to
study the MW NSC in detail are by \citet{Mezger:1999eu},
\citet{Philipp:1999nx}, and \citet{Launhardt:2002nx}. The latter
derive a volume density of $\rho\propto r^{-2}$ for the cluster, with
a core radius of $0.22$\,pc, and estimate its total mass to
$3\pm1.5\times10^{7}$\,M$_{\odot}$.

The center of the Milky Way is the nearest center of a galaxy. From
recent measurements of its distance
\citep{Nishiyama:2006ai,Ghez:2008oq,Groenewegen:2008by,Trippe:2008it,Matsunaga:2009qp,Gillessen:2009qe,Reid:2009nx,Reid:2009eu}
one can calculate a mean value of $R_{0}=8.05\pm0.28$\,kpc. Here, the
unweighted average was computed, with the standard deviation of the
individual measurements as $1\,\sigma$ uncertainty (see also Sch\"odel
et al., 2010, A\&A, in press, arXiv0912.1273). At 8\,kpc, $1"$ of projected distance
corresponds to 0.039\,pc.  This makes the nuclear cluster at the
Galactic center (GC) the only NSC that can be resolved down to scales
of a few milliparsecs with current instrumentation.

\begin{figure}[!htb]
\centering
\includegraphics[width=.7\textwidth]{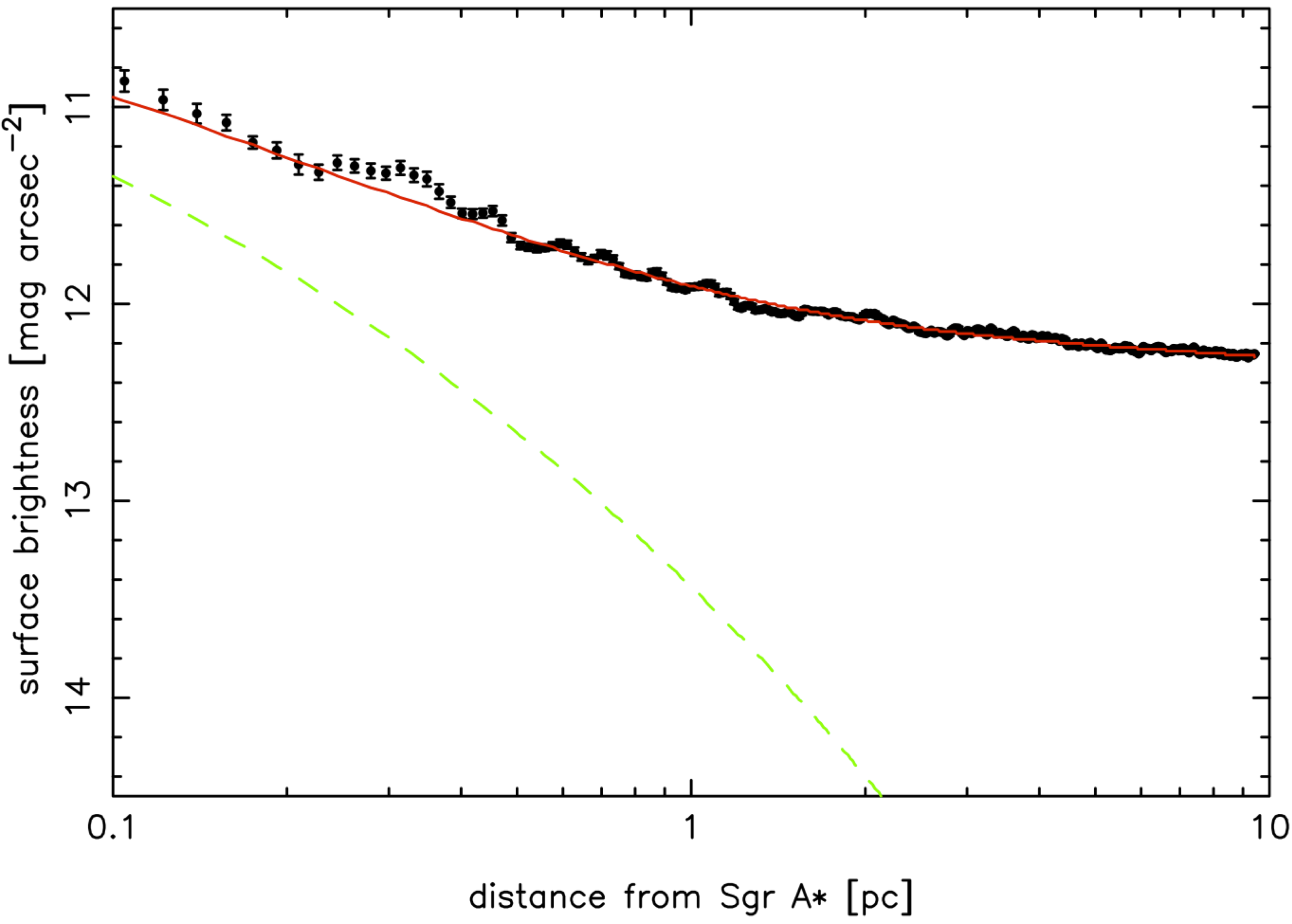}
\caption{\label{Fig:profile} Azimuthally averaged light density profile of the Milky Way nuclear
  star cluster, measured on the SIRIUS/IRSF $Ks$-band image, shown in
  Fig.\,\ref{Fig:MWNSC}. The straight line is a fit with a S\'ersic
  model (index $\sim3.4$) plus a (constant) contribution from the
  nuclear bulge. The derived half light radius is $\sim5$\,pc. The
  dashed line shows the profile of the S\'ersic model cluster.}
\end{figure}

As can be seen in Fig.\,\ref{Fig:MWNSC}, strong and variable
extinction poses serious obstacles to studying the exact shape and
size of the Milky Way's central star cluster. \citet{Graham:2009lh}
model the MW NSC with a S\'ersic function with an index of $\sim3$ and
obtain an effective half-light radius $R_{e}=3.2$\,pc.  They used the
NSC light profile determined from 2MASS data by
\citet{Schodel:2008uq}. Because of saturation issues, 2MASS data are
not ideal for examining the GC.  Using the $Ks$-band image from the
IRSF/SIRIUS GC survey \citep[ Fig.\,\ref{Fig:MWNSC}, see,
e.g.,][]{Nishiyama:2006tx}, we have re-determined and calibrated the
light profile of the cluster (Fig.\,\ref{Fig:profile}), assuming
spherical symmetry and neglecting the influence of extinction. A
S\'ersic profile was fitted to the profile and results in a half-light
radius of $\sim5$\,pc, somewhat larger than the one given by
\citet{Graham:2009lh}. This discrepancy is indicative of the
uncertainty of our current knowledge of this value. In this case, the
uncertainty may be largely due to the different data used for
determining the light profile and/or issues of photometric
calibration/offsets. It is clear that more work is needed
here. Particularly, an effective way to take extinction into account
would be of great help. When estimating the mass of the NSC from its
size and luminosity, major sources of uncertainties are the NIR
mass-to-light ratio. One must also be careful to take into account
that the light from the NSC is dominated by a small number of massive
stars \citep[see][]{Launhardt:2002nx,Schodel:2007tw}.

\section{Kinematics of the Milky Way NSC \label{sec:kinematics}}

The kinematics of the MW NSC on large scales was first studied with
spectroscopic measurements
\citep[e.g.,][]{McGinn:1989kx,Sellgren:1990cl}. Indications were found
for an overall rotation of the cluster parallel to Galactic
rotation. This finding was recently confirmed and significantly
improved by the study of \citet{Trippe:2008it} \citep*[see
also][]{Schodel:2009zr}. Proper motion studies of the GC were
initially limited to the central arcseconds around Sgr\,A*
\citep[e.g.,][]{genzel:1997dp,ghez:1998ad}. Only recently, two
independent works measured stellar proper motions out to a projected
distance of $\sim1$\,pc from Sgr\,A* \citep{Trippe:2008it,Schodel:2009zr}.

\begin{figure}[!htb]
\centering
\includegraphics[width=.7\textwidth]{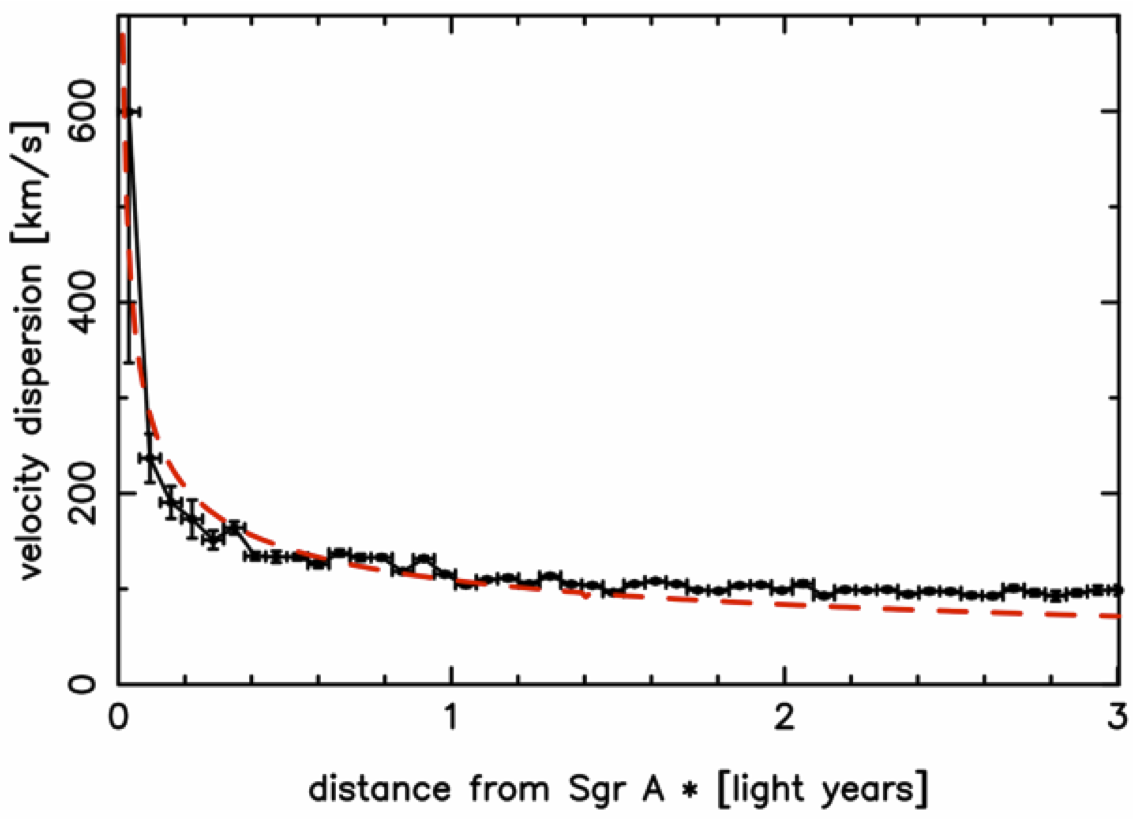}
\caption{\label{Fig:sigma} Azimuthally averaged profile of the
  one-dimensional proper motion velocity dispersion at the GC. The
  dashed line illustrates the velocity dispersion profile of a purely
  Keplerian cluster.}
\end{figure}

Figure\,\ref{Fig:sigma} shows the azimuthally averaged profile of the
one-dimensional proper motion velocity dispersion at the GC, based on
the data from \citet{Schodel:2009zr}. The dashed line indicates the
velocity dispersion profile of a purely Keplerian cluster, projected
onto the plane of the sky. The Keplerian profile is purely
illustrative and no mathematical fit. A mathematical fit is difficult
to achieve because of our insufficient knowledge of the three-dimensional
cluster volume density law of the NSC, particularly in the in the
innermost arcseconds \citep[see discussion
in][]{Schodel:2009zr}. Because of the rotation of the cluster and
possibly present anisotropies, it is not rigorously right to present a
one-dimensional profile of the velocity dispersion
\citep[see][]{Trippe:2008it,Schodel:2009zr}. However, we believe that
Fig.\,\ref{Fig:sigma} serves well to illustrate two points. One of
them is that the proper motions only show a clear Keplerian increase
within $8"$ of the MBH Sgr\,A*, demonstrating the difficulty of
unambiguously proving the existence of a black hole embedded in a
dense stellar cluster, particularly in extragalactic targets. The
other point is that the Kepler law does not fit the data at projected
distances beyond $\sim0.5$\,pc. This is a sign that the gravitational
potential of the NSC itself starts becoming of increasing importance
with distance from the MBH, Sgr\,A*.

Both \citet{Trippe:2008it} and \citet{Schodel:2009zr} used their
proper motion data to estimate the enclosed mass at the GC that is not
contained in Sgr\,A*, i.e. mainly due to stars or stellar
remnants. There are significant differences in the methodology applied
by the two authors. Particularly, the parameterized model of
\citet{Trippe:2008it} takes cluster rotation into account but does not
represent a self-consistent model and does not allow to derive unique
estimates of the enclosed mass \citep[see discussion
in][]{Schodel:2009zr}. The analysis in \citet{Schodel:2009zr} neglects
rotation (which introduces only a small error) and does not make any
assumption about the distribution of the extended (stellar) mass
within 1\,pc of Sgr\,A* other than that it must follow a
power-law. Surprisingly, the latter authors find that even a mass density that
\emph{decreases} toward the black hole can fit the data. This is an
unexpected result and highlights the need for more research,
particularly for better constraints on the three-dimensional volume
density model of the cluster. Under the assumption that the mass
density does in fact \emph{increase} toward Sgr\,A* (as is generally
assumed), the data require at least $0.5\times10^{6}$\,M$_{\odot}$ of
extended mass within 1\,pc of the black hole. If the
mass-to-luminosity ratio is constant in the central parsec, there must
be as much as $1.5\times10^{6}$\,M$_{\odot}$ of extended mass. These
values do not require any special assumptions on the composition of
the cluster.

\section{Is there a cusp around Sgr\,A*?}

Numerous studies on the large-scale structure of the stellar cluster
at the GC have found a volume density that depends on the distance
from Sgr\,A* roughly as $r^{-2}$ \citep[see introduction and
discussion in][]{Schodel:2007tw}. Note, however, that the S\'ersic
model derived by \citet{Graham:2009lh} results in a steeper power
law. The authors point out that previous works may have been biased by
not having accurately taken into account the contribution from the
nuclear bulge.

Theoretical considerations on stellar dynamics predict the existence
of a \emph{cusp} in the central parts of a dense, dynamically relaxed
star cluster surrounding an MBH \citep[for a reviews
see][]{Merritt:2006ys,Alexander:2007vn}, with volume density laws
between $\rho\propto r^{-1.5\ldots-1.75}$ \citep[in case of strong
mass segregation and for the most massive stars $\rho\propto
r^{-2\ldots-2.75}$,][]{Alexander:2009gd}. Early work on the stellar
number density around Sgr\,A* found an almost flat cluster core, with
a radius $R_{\mathrm core}\approx0.2-0.3$\,pc
\citep[e.g.,][]{Eckart:1993zr,Haller:1996kl,Launhardt:2002nx}. However,
these early studies were seriously limited by incompleteness due to
stellar crowding. \citet{Genzel:2003it} determined the number density
of the NSC with the significantly improved resolution and sensitivity
of adaptive optics observations at an 8m-class telescope and found
that the volume density of stars increases towards Sgr\,A* as
$r^{-1.3\ldots-1.4}$ at distances less than $0.35$\,pc from the black
hole. \citet{Schodel:2007tw} revisited the problem with improved data
and methodology and found that the power-law index of the volume
density of stars near Sgr\,A* is as low as $1.2\pm0.05$. Hence, the
power-law index of the stellar volume density is significantly lower
than predicted by classical cusp formation theories. This indicates
that some of the assumptions on which the theoretical work is based
may not be valid in the case of the GC.

At least the assumption about the cluster being old and dynamically
relaxed appears to be violated to some degree because the MW NSC shows
clear signs of repeated bursts of star formation, with the most recent
one having occurred only a few Myr ago
\citep[e.g.,][]{Krabbe:1995fk,Paumard:2006xd,Maness:2007sj}. The
timescale for two-body relaxation in the central parsec of the GC,
however, is generally assumed to be $>1$\,Gyr
\citep[see][]{Alexander:2007vn,Merritt:2009kx}. Moreover,
\citet{Genzel:2003it} showed that the fraction of young, massive stars
increases toward Sgr\,A*. Young stars dominate the number counts in
the central arcseconds \citep[see also][]{Eisenhauer:2005vl}.

\begin{figure}[!htb]
\centering
\includegraphics[width=.7\textwidth]{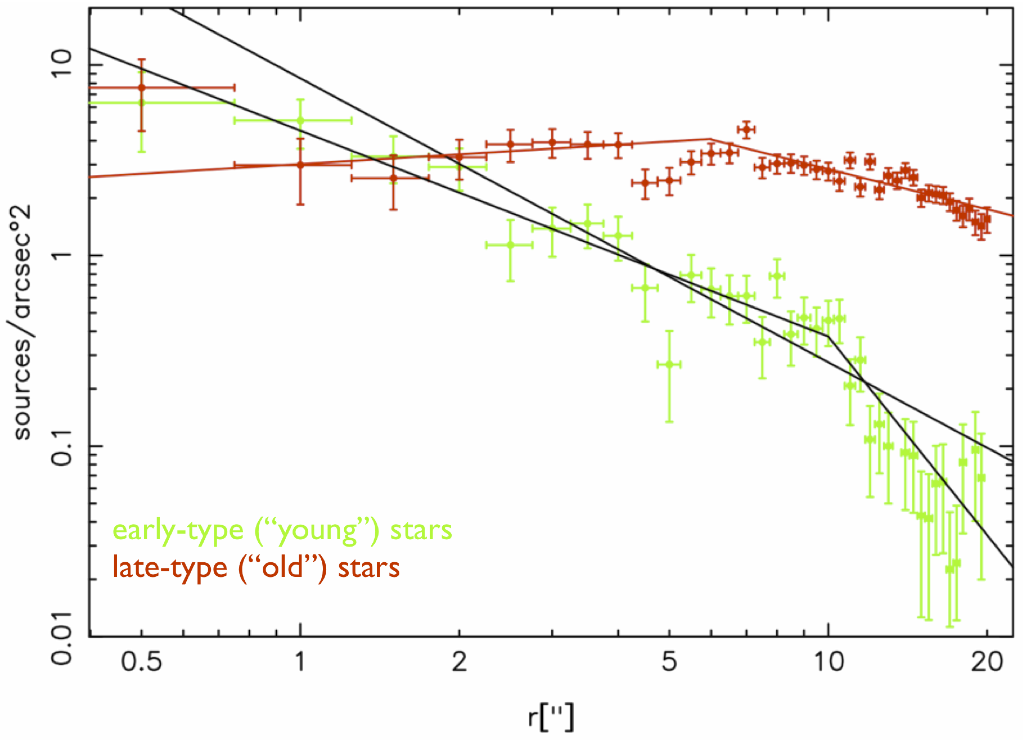}
\caption{\label{Fig:buchholz} Projected surface density of late-type (red/dark grey) and
  early-type (green/light grey, lower counts) stars according to
  \citep{Buchholz:2009fk}. The straight lines are fits with single or
  broken power-laws.}
\end{figure}

Recently, \citet{Buchholz:2009fk} have photometrically classified
stars down to $Ks\approx15.5$ within 1\,pc of Sgr\,A*. They could thus
determine the stellar surface density for early- and late-type stars
separately (see Fig.\,\ref{Fig:buchholz}). The vast majority of the
late-type stars in their sample is expected to be older than about
1\,Gyr and therefore possibly dynamically
relaxed. \citet{Buchholz:2009fk} found that the surface density of
late-type stars is constant within $\sim10"$ of Sgr\,A* and may be
even slightly decreasing in the innermost arcseconds. While a lack of
giant stars in the innermost arcseconds had been noted earlier
\citep[e.g.,][]{Genzel:1996ee,Haller:1996kl}, those early observations
were limited to the brightest giants ($Ks\lesssim12$). The findings of
\citet{Buchholz:2009fk} were confirmed by spectroscopic studies, which
cover, however, smaller areas of the NSC and are probably slightly
shallower \citep{Do:2009tg,Bartko:2010ly}. \citet{Do:2009tg} show that
if the de-projected density profile for the late-type stars is given
by $n(r)\propto r^{-\gamma}$, then $\gamma<1.0$ at the $99.7\%$
confidence level. This is in clear contradiction to the expected
presence of a stellar cusp around Sgr\,A*.

An important caveat that has to be taken into account at this point is
that because of the extreme stellar crowding at the GC, current
instrumentation only detects the brightest stars. The completeness
limit of current imaging observations is $Ks\approx18$, which
corresponds to a mean stellar mass of about 2\,M$_{\odot}$ \citep[see
Fig.\,16 in][]{Schodel:2007tw}. We are only observing the tip of the
iceberg. Nevertheless, lighter stars should be pushed outward by the
heavier components of the cluster and are not expected to show a
steeper density law than the visible stars. Finally, there may be a
cusp formed by stellar mass BHs around Sgr\,A*
\citep{Morris:1993ve,Miralda-Escude:2000qf}. Observational proof of
such a hypothetical cluster of stellar BHs is very difficult to
obtain, but the statistics of detected X-ray transients in the GC may
indicate such a concentration of stellar BHs in the central parsec
\citep{Muno:2005bh}.

Currently, there is no satisfactory explanation for the absence of a
cusp in the observed stars around Sgr\,A*. Working hypotheses include
the collisional destruction of the envelopes of giant stars in the
densest parts of the cluster
\citep[e.g.,][]{Freitag:2008nx,Dale:2009ca} or the infall of an
intermediate mass black hole into the GC, which may have destroyed the
cusp\citep[see][]{Merritt:2006dq}.

\section{Summary}

\begin{figure}[!htb]
\centering
\includegraphics[width=.8\textwidth]{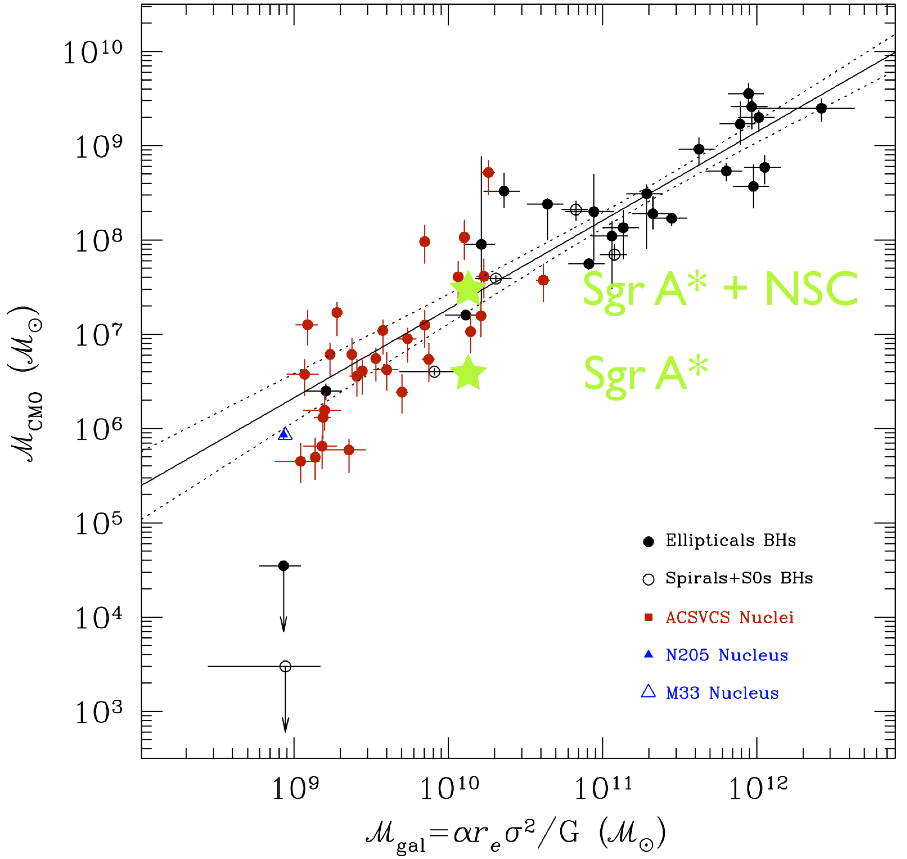}
\caption{\label{Fig:ferrarese} Mass of SMBHs (black
  dots) and nuclear star clusters (grey dots) vs.\ galaxy (spheroid)
  mass. The image has been adapted   from
 Fig.\,2 of   \citet{Ferrarese:2006ly} and is reproduced by permission
 of the AAS. \emph{CMO} stands for \emph{central
    massive object}. See their Fig.\,2 for more details. The lower
  asterisk indicates the mass of Sgr\,A* alone, the upper one the mass
  of Sgr\,A* plus the mass of the Milky Way NSC. The Milky Way bulge
  mass is taken from \citet{Dwek:1995uq} and \citet{Cardone:2005fk}.}
\end{figure}

Nuclear star clusters are found at the dynamical centers of the
majority of spiral, dwarf ellipticals (spheroidal galaxies), and S0 galaxies. With effective
radii of a few pc and masses of $10^{6}-10^{7}$\,M$_{\odot}$ they are
probably the densest and most massive star clusters in the
Universe. They show a complex stellar population, produced by repeated
star formation events. Intriguingly, there appears to exist a similar
relation between the mass of NSCs and the mass of the host galaxy
(respectively its spheroid) as between SMBHs and
the host galaxies (see Fig.\,\ref{Fig:ferrarese}). This may indicate
that there is always a \emph{central massive object} (CMO) present at
galaxy centers, either preferentially in the form of an NSC in
low-mass galaxies or in the form of an SMBH in the
most massive galaxies
\citep[e.g.,][]{Ferrarese:2006ly,Graham:2009lh}. Massive black holes
may well be present at the nuclei of almost all galaxies, but it can
be extremely difficult to prove unambiguously the presence of a black
hole inside a NSC (see section\,\ref{sec:kinematics}). In the
discussion about CMOs it is important to point out that the relation
between massive black holes and NSCs is not clear at this point. For
example, \citet{Kormendy:2009fk} argue that this apparent relation
could be an accident and that there is no further relation between
MBHs and NSCs than that they are both probably fed by gas from the
surrounding galactic disk. Additional reliable measurements of
extragalactic MBH and NSC masses are needed  to
resolve this issue.

The cluster at the center of the Milky Way appears to be very similar
to its extragalactic cousins and thus represent a valid template for
their understanding. It has a mass of one to a few
$10^{7}$\,M$_{\odot}$, an effective radius of 3-5\,pc, and shows a
complex star formation history. Adding the mass of the MW NSC to the
mass of the central black hole, Sgr\,A*, makes the central massive
object at the GC fit much better on the relationship between CMO and
galaxy mass than if only considering the BH mass alone
(Fig.\,\ref{Fig:ferrarese}). Research questions with respect to the
Galaxy's nuclear star cluster that should be addressed in the coming
years are: How can extinction be taken into account when studying the
MW NSC? Is it really spherically symmetric? What mathematical model
fits best its shape and what is its correct effective radius? How is
the extended mass distributed in the central parsec? Is there a hidden
cusp of stellar mass black holes? What is the best  explanation for
the absence of an observed cusp around Sgr\,A* in the old stellar population?

\acknowledgements I am grateful to S. Nishiyama for providing the
image of the Milky Way nuclear star cluster from the IRSF/SIRIUS survey.

\bibliography{/Users/rainer/Documents/Papers/BibDesk/BibGC}

\end{document}